# A metamaterial-free fluid-flow cloak

Fuyang Tay[1,#], Youming Zhang[1,#], Hongyi Xu[1,#], Honghui Goh[1], Yu Luo[2,(a)], Baile Zhang[1,3,(b)]

[1]*Division of Physics and Applied Physics, School of Physical and Mathematical Sciences, Nanyang Technological University, Singapore 637371, Singapore*

[2]*School of Electrical and Electronic Engineering, Nanyang Technological University, Singapore 639798, Singapore*

[3]*Centre for Disruptive Photonic Technologies, Nanyang Technological University, Singapore 637371, Singapore*

# *These authors contributed equally to this work*

(a) E-mail: luoyu@ntu.edu.sg

(b) E-mail: blzhang@ntu.edu.sg

## Abstract

The model of ideal fluid flow around a cylindrical obstacle exhibits a long-established physical picture where originally straight streamlines will be deflected over the whole space by the obstacle. As inspired by transformation optics and metamaterials, recent theories have proposed the concept of fluid cloaking able to recover the straight streamlines as if the obstacle does not exist. However, such a cloak, similar to all previous transformation-optics-based devices, relies on complex metamaterials, being difficult to implement. Here we deploy the theory of scattering cancellation and report on the experimental realization of a fluid-flow cloak without metamaterials. This cloak is realized by engineering the geometry of the fluid channel, which effectively cancels the dipole-like scattering of the obstacle. The cloaking effect is demonstrated via direct observation of the recovered straight streamlines in the fluid flow with injected dyes. Our work sheds new light on conventional fluid control and may find applications in microfluidic devices.

Ideal fluid flow around a cylinder is a fundamental problem discussed in many textbooks of fluid mechanics [1-4]. Being inviscid and incompressible, an ideal fluid satisfies the mass continuity equation, which can be simplified into Laplace's equation at steady states. When encountering a circular cylinder, the ideal fluid no longer follows straight streamlines, but flows around the cylinder with deflected streamlines described by a conformal mapping [1, 4]. This model provides a physical picture to understand general fluid flow in fluid mechanics when a complex-shaped obstacle or fluid viscosity is involved.

Recently, with the inspiring development of transformation optics and metamaterials, substantial interests arose in constructing invisibility cloaking devices to hide an object from external physical fields [5-25]. This kind of transformation-optics cloaking devices were firstly proposed and realized for electromagnetic waves [5-16], and then immediately extended to acoustic waves [17-20], heat flow [21-24], and other field forms [25]. In 2011, this transformation-optics cloaking concept has been further extended to fluid flow [26]. However, similar to all previous transformation-optical devices, this fluid-flow cloak has spatially variant material parameters and thus requires complex metamaterials, being difficult to realize [26-28]. On the other hand, scattering cancellation is another powerful approach for cloak design [29-35]. It was originally proposed for plasmonic particles in quasistatic electric fields [29, 30], and recently has been extended into magnetic fields [31, 32], heat conduction [33, 34], and direct currents [35]. However, its application in fluid control has never been discussed.

In this Letter, we extend the previous scattering cancellation approach to fluid control, and construct a fluid-flow cloak to hide a cylindrical obstacle without disturbing the straight external streamlines (see the movie in Supplementary Material demonstrating the effectiveness of such a fluid cloak [36]). In particular, the use of scattering cancellation in fluid flow has gained an unprecedented unique feature, i.e., being "metamaterial-free": our fluid cloak is realized by changing the geometry of the fluid channel, rather than employing any complex metamaterial design. By injecting dye particles into the fluid flow, we have observed the successful recovery of straight streamlines passing through the obstacle, as if the obstacle does not exist.

We shall firstly point out that the ideal fluid with zero viscosity in fact does not exist in nature (it is "dry water" as quoted from John von Neumann [37]). The flow of a real fluid with finite viscosity is governed by the Navier-Stokes equation, which has nearly no analytical solution due to its nonlinear viscosity term. Nevertheless, a viscous fluid flow in a narrow gap between two parallel plates, known as Hele-Shaw flow, can be described by a scalar potential function, exhibiting similar features of two-dimensional (2D) ideal fluid flow [3, 4, 38, 39]. Moreover, Hele-Shaw flow plays a significant role in microfluidic devices and plastic-forming manufacturing operations. A realistic cloak for Hele-Shaw flow may find useful applications in these related fields.

Let us start with Fig. 1(a), which shows the ideal fluid flow around a cylinder with radius $R_1$ in a 2D geometry [1-4]. By denoting the stream function as $\psi$ and the velocity potential as $\phi$ (here $\psi$ describes the streamlines, and $\nabla \phi$ gives the flow velocity $\bar{v}$; $\psi$ and $\phi$ satisfy Cauchy-Riemann conditions), we can describe the flow with a complex potential $w(z) = \phi(x,y) + i\psi(x,y)$ in the complex plane $z = x + iy$. Without the cylinder, the original uniform flow with straight streamlines exhibits the complex potential $w = Uz$, where $U$ is the speed of the stream that flows uniformly in the $\hat{x}$ direction. The presence of the cylinder deflects the flow according to the conformal mapping of $z \to z + R_1^2/z$ [1-4]. Thus the complex potential $w$ is mapped to $w' = U(z + R_1^2/z)$ in the region outside of the cylinder (i.e., $|z| > R_1$). Note that the newly produced term $UR_1^2/z$ corresponds to the complex potential of a "dipole"-like doublet (i.e., a point source and a point sink placed extremely close to each other, similar to a dipole of positive charge and negative charge in electromagnetics) which has the dipole strength vector $-2\pi U R_1^2 \hat{x}$. Therefore, in the language of electromagnetics, the cylinder induces a "dipole field" of fluid flux, disturbing the flow at every location.

The analogy to electromagnetics implies that it might be straightforward to construct a fluid cloak by applying the scattering cancellation. Indeed, we can consider a cloak as shown in Fig. 1(b), which consists of a shell with an outer radius $R_2$ and an inner radius $R_1$ that encloses the cylindrical obstacle completely. This cloak can guide the fluid flux smoothly around the obstacle leaving the external fluid flux undisturbed. In the previous magnetic/thermal cloaks [31, 33, 34], the 2D calculation from Laplace's equation requires their magnetic permeability/thermal conductivity to take the relative value of $(R_2^2 + R_1^2)/(R_2^2 - R_1^2)$. By the same token, we can simply write down the mathematical condition of the fluid cloak as [36]

$$\rho_2 = \frac{R_2^2 + R_1^2}{R_2^2 - R_1^2}\rho_1. \#(1)$$

Here $\rho_1$ is the fluid density in the background, and $\rho_2$ is that inside the cloak shell.

However, we still have two problems to tackle. Firstly, Eq. (1) requires the fluid density inside the cloak to be compressed compared to that in the background. This contradicts with the incompressibility of the ideal fluid as well as most real fluids in general. From the viewpoint of practical implementation, any solid material implemented in the cloak shell will block the fluid flow and therefore effectively dilute the local fluid density. Secondly, Eq. (1) is derived based on Laplace's equation which applies only to the ideal fluid that has negligible viscosity. However, a real fluid must contain finite viscosity (an extremely low viscosity in a real fluid also comes along with an extremely high Reynold number (Re $\gg$ 1), marking the onset of turbulence). The influence of the viscosity remains an issue to the cloaking condition in Eq. (1).

To tackle the problems mentioned above, we consider the creeping flow with low Reynold number (Re $\ll$ 1) in a narrow gap between two plates, which is known as Hele-Shaw flow. In this case, a viscous flow can be simplified into an ideal fluid flow satisfying Laplace's equation [3, 4, 38, 39]. Hence, the problem of viscosity can be solved. As illustrated in Fig. 1(c), we consider fluid with a density $\rho_1$ that flows into a narrow channel with a height of $h_b$. A solid cylinder with radius $R_1$ that penetrates through the channel serves as a cylindrical obstacle. So, it can be expected that the viscous flow in the narrow channel in the presence of the cylindrical obstacle will behave like the picture in Fig. 1(a) (there will be some discrepancies in thin layers close to the boundary of the obstacle; to be discussed later).

Now we design the cloak. As mentioned above, it is impractical to compress the fluid density to fulfill the cloaking condition in Eq. (1). However, from the 2D perspective, we can emulate a higher local fluid density $\rho_2$ by extending the height $h_s$ of the channel at the cloak shell region, as illustrated in Fig. 1(c). According to mass conservation, the extended height required to construct a fluid flow cloak satisfies the formula below,

$$h_\text{s} = \frac{R_2^2 + R_1^2}{R_2^2 - R_1^2} h_\text{b} \quad \#(2)$$

It is worth mentioning that the value of $h_\text{s}$ in Eq. (2) is only an estimation. The no-slip condition (i.e., only zero velocity is allowed at the boundary) gives rise to distortion of streamlines in the Hele-Shaw flow in the vicinity of boundaries of the obstacle, being different from the ideal fluid flow. Hence, the optimal height for the fluid flow cloak should be slightly shifted from $h_\text{s}$ directly calculated from Eq. (2). The method we used to optimize $h_\text{s}$ will be discussed in a later section.

Figure 2 shows the schematic diagram of the experimental setup. We designed a rectangular channel with dimensions 146 mm × 50 mm × 5 mm. A cylindrical obstacle with a radius $R_1 = 8$ mm was placed at the center of the channel. As the fluid is affected by gravity, we extruded the cloak shell region with an outer radius $R_2 = 14$ mm along $-\hat{z}$ direction, like a trough surrounding the obstacle. The height of the shell region measured from the top of the flowing channel to the bottom of the trough is represented by $h_\text{s}$. During the experiment, an electrically driven piston pump was used to pump the fluid into the setup through a thick rubber tube. The fluid first filled up a sink, and then flowed into the channel uniformly. Glycerin was used in our demonstration due to its high viscosity (about 0.63 Pa · s for 95% glycerin solution at room temperature) [40]. The Reynold number, which is required to be smaller than 1 for the creeping flow, is about 0.03 in our case (see the calculations in Supplementary Material) [36]. Moreover, the diameter of the obstacle $2R_1 = 16$ mm is larger than the gap of the channel $h_\text{b} = 5$ mm, which is consistent with the Hele-Shaw flow [3, 4, 38, 39]. Three samples were prepared, (i) a reference sample without obstacle; (ii) an obstacle sample without cloak; and (iii) an obstacle sample with the cloak. They were manufactured through 3D printing. A piece of glass block was placed on top of the samples to enclose the flow channel. The indicators, the glycerin dyed with black acrylic paints, were injected manually through four evenly distributed thin rubber tubes when the system became steady to visualize the streamlines. Apart from that, the background fluid was dyed with white acrylic paints to improve the visualization of the streamlines. The experiments were recorded by a camera from top.

Here we introduce our optimization method for the height of the cloak shell. The initial value of $h_\text{s} = 9.85$ mm is obtained from Eq. (2). Therefore, the simulation using the same setup with the experiment was repeated with a range of $h_s$ close to 9.85 mm in COMSOL Multiphysics 5.2 in order to obtain the optimized $h_s$. The creeping flow that is governed by Stokes equations was used in the simulation. It is worth noting that the optimized $h_s$ should show straight streamlines in the background region, as shown in Fig. 1(b). Hence, we first extracted the $y$-coordinates of 40 streamlines, which were spaced equally in the $y$-direction, from each simulation. Next, we removed those $y$-coordinates that were far away from the obstacle or located inside the cloak shell region (see Fig. S1 in Supplementary Material [36]). The standard deviation of remaining $y$-coordinates extracted from each streamline was calculated individually, and finally we defined $y$-variation as the mean of those standard deviations. The dependence of $y$-variance on $h_s$ is illustrated in Fig. 3, and the optimized $h_s$, which should give the smallest $y$-variance, is shown to be 10 mm. The inset demonstrates the simulation result when $h_s$ = 10 mm. The color represents the magnitude of total velocity. The lines and arrows with teal color denote the streamlines and direction of flowing velocity, respectively. As anticipated, the streamlines are almost undisturbed in the background region. The flowing velocity is lower (represented by deeper color) near to the wall and the obstacle due to the boundary effect.

The movie provided in Supplementary Material [36] recorded the dynamic process of the fluid flow passing by the obstacle. We extracted the snapshots at 4 seconds and 10 seconds of the video for illustration in Fig. 4. As mentioned before, the experiments were repeated with three samples and the glycerin was input from the top side of the snapshots. Four streamlines were visualized by the black indicators and they were denoted by indices, "1", "2", "3" and "4" [Fig. 4(a)].

Figure 4(a) and (b) verify that the flow generated by the pump was uniform and the streamlines were straight if no obstacle existed. The black shadows presented in Fig. 4(a) and (b) were just the reflections of the camera lens by the glass block placed on top of the sample. In contrast, in the presence of the cylindrical obstacle, the fluid flow was blocked by the obstacle and the streamlines were deflected to the left and right sides of the obstacle [Fig. 4(c) and (d)]. Although the side walls in our experiment might slightly reduce the distortion of streamlines in its vicinity, Fig. 4(d) still showed a similar pattern to Fig. 1(a). It is worthwhile noting that the distortion of streamlines arose even in places away from the obstacle.

Finally, the results for the obstacle sample with the cloak, which demonstrates the realization of fluid cloaking, are as shown in Fig. 4(e) and (f). The cloaking shell region was represented by a pink dotted circle. As anticipated, the distortion of streamlines outside the shell region was cancelled and the streamlines were straight. The background region in Fig. 4(f) exhibited the same pattern as shown in Fig. 4(b) and 1(b). The cylindrical obstacle was therefore "hidden" from the external fluid fluxes.

In summary, our work demonstrates an innovative approach to realize a fluid flow cloak that can hide a cylindrical obstacle in a narrow fluid channel. In contrast to previous designs employing complex metamaterials, our fluid cloak adopts the scattering cancellation approach and is realized by merely adjusting the geometry of the fluid channel. Such a simple solution provides a new venue for fluid control by transplanting novel concepts of metamaterials (rather than using complex metamaterials themselves) to fluid engineering.

Note: In the final stage of submission, we found two publications reporting hydrodynamic cloaks [41,42], and one of them implements a fluid flow cloak with multilayer metamaterials [42]. The fundamental difference of our work lies in the scattering cancellation in fluid control and the absence of complex metamaterial design.

**Acknowledgement**
This work was sponsored by Singapore Ministry of Education under Grants No. MOE2018-T2-1-022 (S), MOE2018-T2-2-189 (S), MOE2016-T3-1-006, RG174/16 (S), 2017-T1-001-239 RG91/17-(S).

# References

[1] Y. Nakayama, *Introduction to fluid mechanics* (Butterworth-Heinemann, 1999).
[2] R. W. Fox, A. T. McDonald, and P. Pitchard, *Introduction to Fluid Mechanics* (John Wiley & Sons, Inc, 2004), 6 edn.
[3] Y. A. Cengel, *Fluid mechanics* (Tata McGraw-Hill Education, 2010).
[4] P. K. Kundu, I. M. Cohen, and D. R. Dowling, *Fluid mechanics* (Academic Press, 2016), 6 edn.
[5] U. Leonhardt, Science **312**, 1777 (2006).
[6] J. B. Pendry, D. Schurig, and D. R. Smith, Science **312**, 1780 (2006).
[7] D. Schurig, J. Mock, B. Justice, S. A. Cummer, J. B. Pendry, A. Starr, and D. Smith, Science **314**, 977 (2006).
[8] R. Liu, C. Ji, J. Mock, J. Chin, T. Cui, and D. Smith, Science **323**, 366 (2009).
[9] J. Valentine, J. Li, T. Zentgraf, G. Bartal, and X. Zhang, Nat. Mater. **8**, 568 (2009).
[10] L. H. Gabrielli, J. Cardenas, C. B. Poitras, and M. Lipson, Nat. Photonics. **3**, 461 (2009).
[11] H. F. Ma and T. J. Cui, Nat. Commun. **1**, 21 (2010).
[12] T. Ergin, N. Stenger, P. Brenner, J. B. Pendry, and M. Wegener, Science **328**, 337 (2010).
[13] X. Chen, Y. Luo, J. Zhang, K. Jiang, J. B. Pendry, and S. Zhang, Nat. Commun. **2**, 176 (2011).
[14] B. Zhang, Y. Luo, X. Liu, and G. Barbastathis, Phys. Rev. Lett. **106**, 033901 (2011).
[15] N. Landy and D. R. Smith, Nat. Mater. **12**, 25 (2013).
[16] X. Ni, Z. J. Wong, M. Mrejen, Y. Wang, and X. Zhang, Science **349**, 1310 (2015).
[17] M. Farhat, S. Enoch, S. Guenneau, and A. Movchan, Phys. Rev. Lett. **101**, 134501 (2008).
[18] B.-I. Popa, L. Zigoneanu, and S. A. Cummer, Phys. Rev. Lett. **106**, 253901 (2011).
[19] S. Zhang, C. Xia, and N. Fang, Phys. Rev. Lett. **106**, 024301 (2011).
[20] L. Zigoneanu, B.-I. Popa, and S. A. Cummer, Nat. Mater. **13**, 352 (2014).
[21] C. Fan, Y. Gao, and J. Huang, Appl. Phys. Lett. **92**, 251907 (2008).
[22] S. Guenneau, C. Amra, and D. Veynante, Opt. Express **20**, 8207 (2012).
[23] S. Narayana and Y. Sato, Phys. Rev. Lett. **108**, 214303 (2012).
[24] R. Schittny, M. Kadic, S. Guenneau, and M. Wegener, Phys. Rev. Lett. **110**, 195901 (2013).
[25] Q. Ma, Z. L. Mei, S. K. Zhu, T. Y. Jin, and T. J. Cui, Phys. Rev. Lett. **111**, 173901 (2013).
[26] Y. A. Urzhumov and D. R. Smith, Phys. Rev. Lett. **107**, 074501 (2011).
[27] Y. A. Urzhumov and D. R. Smith, Phys. Rev. E **86**, 056313 (2012).
[28] D. Culver and Y. Urzhumov, Phys. Rev. E **96**, 063107 (2017).
[29] A. Alù and N. Engheta, Phys. Rev. E **72**, 016623 (2005).
[30] B. Edwards, A. Alù, M. G. Silveirinha, and N. Engheta, Phys. Rev. Lett. **103**, 153901 (2009).
[31] F. Gömöry, M. Solovyov, J. Šouc, C. Navau, J. Prat-Camps, and A. Sanchez, Science **335**, 1466 (2012).
[32] S. Narayana and Y. Sato, Adv. Mater. **24**, 71 (2012).
[33] H. Xu, X. Shi, F. Gao, H. Sun, and B. Zhang, Phys. Rev. Lett. **112**, 054301 (2014).
[34] T. Han, X. Bai, D. Gao, J. T. Thong, B. Li, and C.-W. Qiu, Phys. Rev. Lett. **112**, 054302 (2014).
[35] T. Han, H. Ye, Y. Luo, S. P. Yeo, J. Teng, S. Zhang, C.-W. Qiu, Advanced Materials **26**, 3478 (2014).
[36] See Supplemental Material at [URL] for the video and calculations.
[37] R. P. Feynman, R. B. Leighton, and M. Sands, *The Feynman lectures on physics, vol. 2* (Addison-Wesley, 1979).
[38] H. S. Hele-Shaw, Nature **58**, 34 (1898).
[39] M. Van Dyke, *An album of fluid motion* (The Parabolic Press, 1982).
[40] D. D. Atapattu, R. P. Chhabra, and P. H. T. Uhlherr, J. Non-Newton Fluid **59**, 245 (1995).
[41] S. Zou, Y. Xu, R. Zatianina, C. Li, X. Liang, L. Zhu, Y. Zhang, G. Liu, Q. H. Liu, H. Chen, and Z. Wang, Phys. Rev. Lett. **123**, 074501 (2019).
[42] J. Park, J. R. Youn, and Y. S. Song, Phys. Rev. Lett. **123**, 074502 (2019).

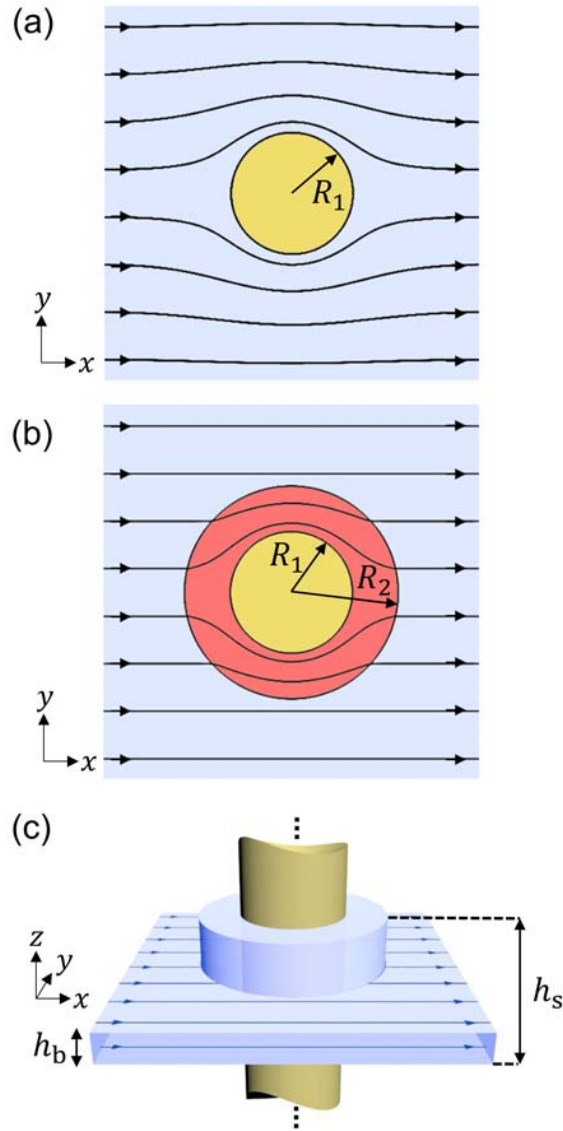

**Fig. 1 Design of a fluid cloak. (a)** Streamlines for 2D ideal fluid flow around a circular cylinder with radius $R_1$. **(b)** A hypothetical fluid cloak that can guide fluid flow around the cylindrical obstacle without disturbing the external straight streamlines. The cloak has an outer radius $R_2$ and inner radius $R_1$. **(c)** Conceptual illustration of the fluid cloak for the flow in a narrow fluid channel with a height $h_b$. The fluid cloak is realized by increasing the height of the cloaking shell region to $h_s$.

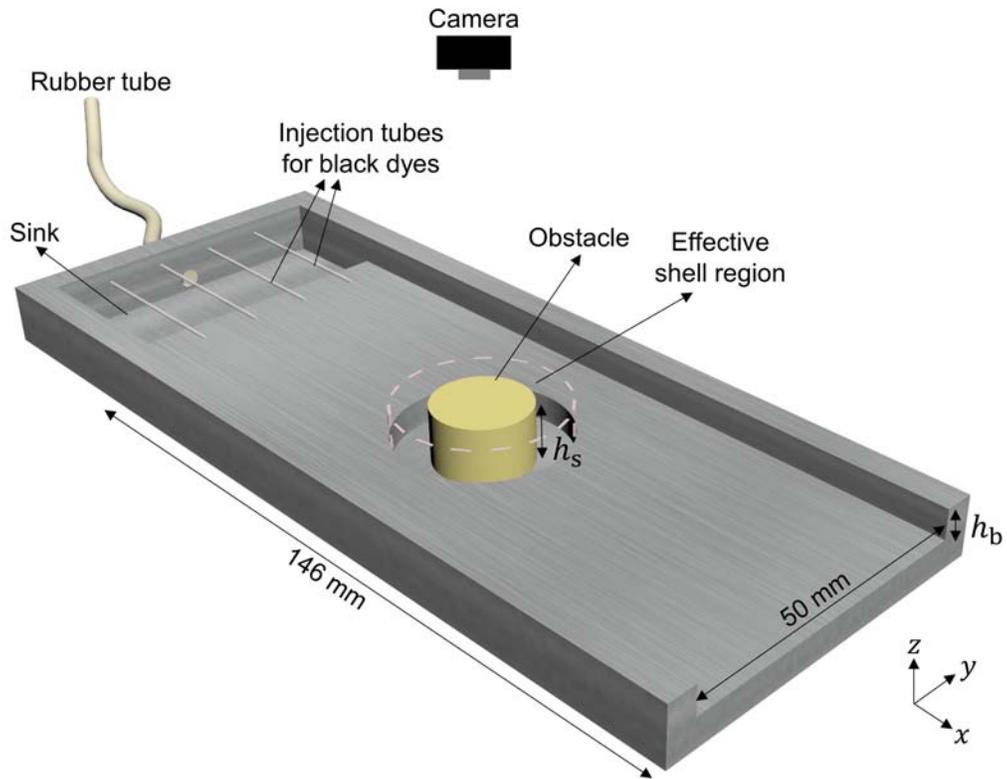

**Fig. 2 Schematic diagram of the experimental setup.** The glycerin that dyed by white paints was pumped into a sink first through a wide rubber tube and then flowed into a rectangular channel uniformly. The indicators, glycerin dyed by black dyes, were injected into the channel manually with injection syringes through the injection tubes at steady state. The flowing channel has a width of 50 mm and a length of 146 mm. A piece of glass block was placed on top of the samples to enclose the channel. The shell region was extended downwards, like a trough surrounding the obstacle. The height of the background and effective shell region are $h_b$ = 5 mm and $h_s$ = 10 mm respectively. The entire processes were recorded by a camera from top.

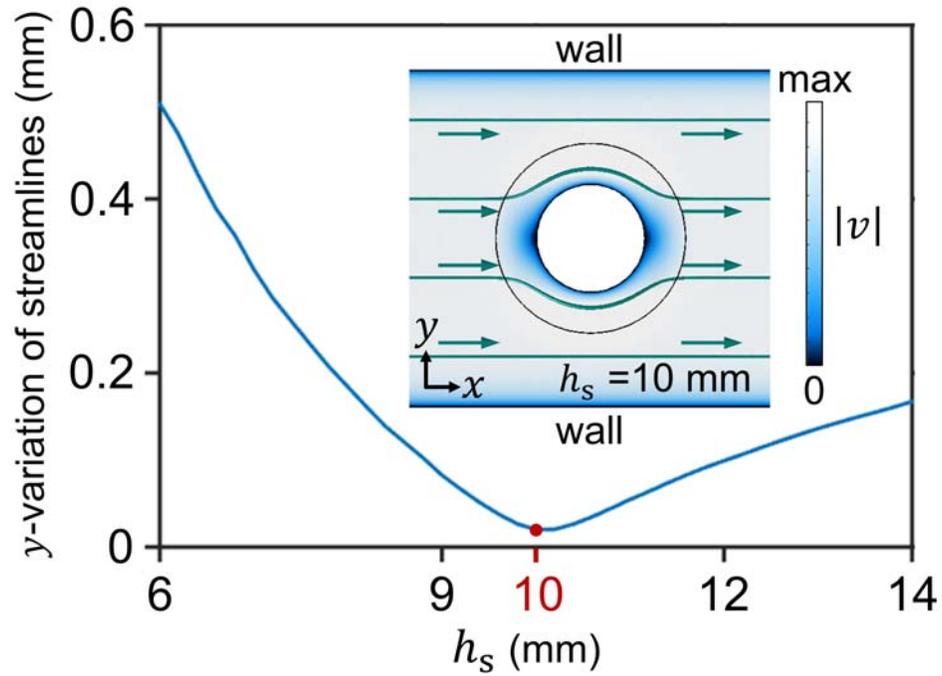

**Fig. 3 Optimization of the cloak geometry.** The simulation using the same setup with the experiment was repeated with different $h_s$ in COMSOL Multiphysics 5.2. The minimum of the defined $y$-variation is obtained when $h_s = 10$ mm (represented by a red dot). The inset shows the simulation result when $h_s = 10$ mm. The color slice represents the magnitude of velocity. The circle with a black solid line shows the cloaking shell. The lines and arrows with teal color represent the streamlines and flowing velocity respectively. The streamlines in the background region remain straight.

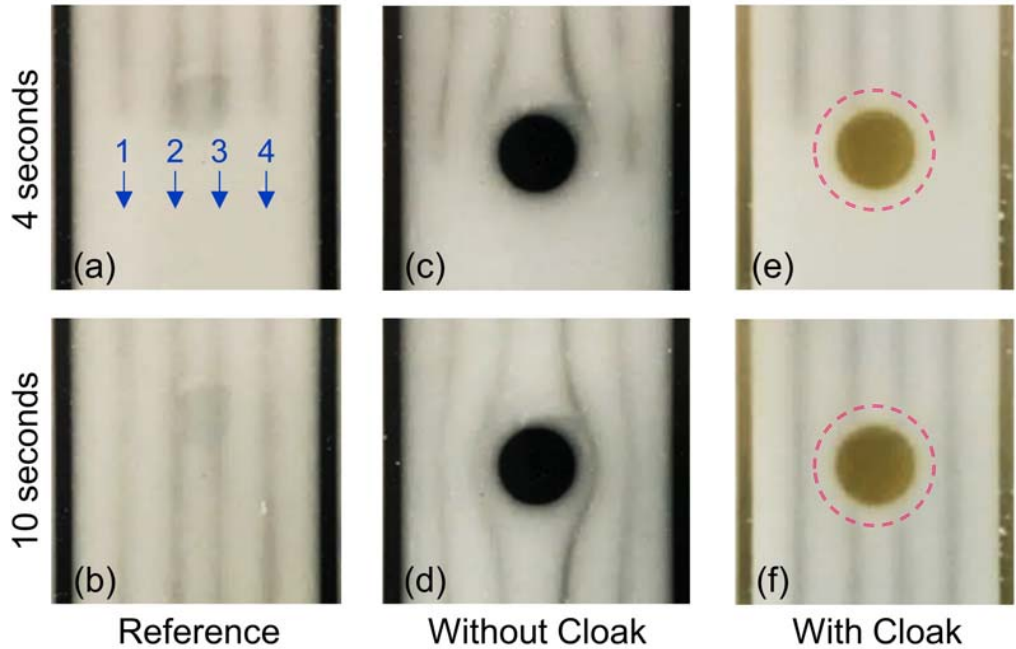

**Fig. 4 Experimental demonstration of the fluid cloak. (a)-(b)** Observed streamlines flowing through a reference sample without the obstacle. The fluid flowed from the top side of the photos. The streamlines are denoted by the indices from 1 to 4. The shadow is the reflection of the camera lens by the glass block on top. **(c)-(d)** Observed streamlines flowing through an obstacle sample without the cloak. The streamlines were deflected by the obstacle. **(e)-(f)** Observed streamlines flowing through an obstacle sample with the cloak. The cloak region is represented by a pink dotted circle. The streamlines outside the cloaking shell region were undisturbed by the obstacle.